\documentstyle[12pt,psfig]{article}

\makeatletter
\def\setcaption#1{\def\@captype{#1}}
\makeatother

\newcommand {\beq}{\begin{equation}}
\newcommand {\eeq}{\end{equation}}
\newcommand {\beqa}{\begin{eqnarray}}
\newcommand {\eeqa}{\end{eqnarray}}
\newcommand {\beqan}{\begin{eqnarray*}}
\newcommand {\eeqan}{\end{eqnarray*}}

\newcommand {\Romannumeral}[1]{\uppercase\expandafter{\romannumeral#1}}

\newcommand {\aaa}{\mbox{\scriptsize A}}

\newcommand {\tr}{\mbox{tr}}

\newcommand {\del}{\partial}

\begin{document}
\setlength{\baselineskip}{7mm} %10mm

\begin{titlepage}
 \renewcommand{\thefootnote}{\fnsymbol{footnote}}
    \begin{normalsize}
     \begin{flushright}
                 DPNU-96-40\\
%                 KEK Preprint 95-??\\
                 August 1996
%~~\\
%~~\\
%~~\\
%~~
     \end{flushright}
    \end{normalsize}
    \begin{Large}
       \vspace{2cm}
       \begin{center}
         {\Large On Existence of Nontrivial Fixed Points \\
in Large $N$ Gauge Theory in More than Four Dimensions} \\
%         {\bf Quantum $R^2$ Gravity in the $2+\epsilon$--Dimensional
%Formalism} \\
       \end{center}
    \end{Large}

  \vspace{1cm}

\begin{center}
           Jun N{\sc ishimura}\footnote
           {E-mail address : nisimura@eken.phys.nagoya-u.ac.jp}\\
      \vspace{1cm}
%      \vspace{5mm}
%        $\ast$ 
{\it Department of Physics, Nagoya University,}\\
               {\it Chikusa-ku, Nagoya 464-01, Japan}\\
      \vspace{2cm}

%\vspace{15mm}

\end{center}
\hspace{7cm}

\begin{abstract}
\noindent 
Inspired by a possible relation between large $N$ gauge theory 
and string theory, we search for nontrivial 
fixed points in large $N$ gauge theory in more than four dimensions.
We study large $N$ gauge theory 
%pure SU($N$) lattice gauge theory in the large $N$ limit 
through Monte Carlo simulation of the twisted Eguchi-Kawai model 
in six dimensions as well as in four dimensions.
The phase diagram of the system 
with the two coupling constants which correspond to 
the standard plaquette action and the adjoint term has been explored.
%Since first order phase transition occurs in both dimensions
%when the standard plaquette action is used, 
%we study the system by adding the adjoint term in the action.
%In four dimensions we find that the line of 
%first order phase transition terminates at a point with 
%a sufficiently large negative coefficient of the adjoint term.
%In six dimensions, on the other hand, we argue, 
%with the aid of the factorization valid in the large $N$ limit, 
%that the line of first order phase 
%transition continues to infinity.
%Implications to string theory is also discussed.
%A possible scenario for the supersymmetric case is briefly discussed. 

~


\end{abstract}

\end{titlepage}
\vfil\eject

%\vspace{1cm}

\setcounter{footnote}{0}

\section{Introduction}
%\setcounter{equation}{0}
%\hspace*{\parindent}

String theory has been considered as a natural candidate for 
the unified theory including gravity. 
Intensive study revealed, however, that nonperturbative 
understanding of its dynamics is essential in 
making any predictions as to our present world.
The double scaling limit of the large $N$ matrix model \cite{double} 
is a rare example 
in which nonperturbative study of string theory has been successful.
Although the development concerning matrix models so far is restricted to 
the space-time dimension less than 1, the above success
shows that nonperturbative study of string theory in physical dimensions 
may be possible through the study of the double scaling limit of 
some kind of large $N$ field theory. 

As such a field theory, we consider large $N$ gauge theory. 
It has been generally believed 
that large $N$ gauge theory is related to 
some kind of string theory, 
since 't Hooft showed within perturbation theory 
that the $1/N$ expansion of U($N$) gauge theory gives an expansion 
in terms of the topology of the Feynman diagrams \cite{tHooft}.
If, in the strong coupling regime, dominant contributions come from 
such diagrams as those with very fine internal lines like fishnets, 
the theory can be viewed as a string theory by identifying the fishnet 
with a continuous worldsheet embedded in the space-time where the gauge 
field lives.
Also in lattice gauge theory, the expectation value of a Wilson loop 
can be expressed, in the strong coupling expansion,
as a sum over random surfaces spanning the loop. 
Although no rigorous connection between large $N$ gauge theory 
and string theory has not yet been established, 
there is a good reason to consider it seriously.

In four dimensions, since 
nonabelian gauge theory is asymptotically 
free, a possible relation to string theory can arise only 
as an effective theory in the infrared region, where 
the coupling constant becomes sufficiently large.
This naturally leads us to consider the theory in more than four dimensions.

The $\epsilon$-expansion of $(4+\epsilon)$-dimensional nonabelian 
gauge theory shows that there exists an ultraviolet
fixed point between the confining phase and the deconfining phase when 
$\epsilon$ is positive. 
If this fixed point survives even in physical dimensions 
such as 5 or 6, there is a possibility of 
constructing a string theory 
by approaching the fixed point as one take the large 
$N$ limit.
Considering that the critical dimension of bosonic string 
theory is 26, 
one may speculate that the fixed point might well exist 
in more than four dimensions.

Since gauge theory is perturbatively unrenormalizable in more than 
four dimensions, the existence of such a fixed point is 
quite nontrivial.
An example is three-dimensional nonlinear sigma model, 
which is perturbatively unrenormalizable and yet 
can be defined on the lattice by taking the continuum limit at the 
second order phase transition point. 
Although this example is rather trivial in the sense that 
the continuum theory belongs to the same universality class as 
three-dimensional $\phi^4$ theory which is superrenormalizable,
it shows that constructive definition of perturbatively unrenormalizable 
theory is possible at least {\it in principle}. 
Constructing quantum gravity within ordinary field theory belongs to 
the same kind of problem and we expect that our study may also be suggestive 
to that direction.

Since we need a nonperturbative approach, we consider lattice gauge theory. 
Monte Carlo simulation of large $N$ gauge theory in a straightforward 
manner is, however, unfeasible due to the huge number of dynamical 
degrees of freedom, which is as large as $N^2 \times D \times L^D$, 
where $L$ is the size of the lattice, 
and $D$  is the dimension of the space-time. 
However, as Eguchi and Kawai showed in the early 80's \cite{EguchiKawai}, 
the factorization valid in large $N$ gauge theory leads to the fact that 
the above system is equivalent to a one-site model ($L$=1) with 
$N^2 \times D$ dynamical degrees of freedom. 
This tremendous reduction of dynamical degrees of freedom 
makes the study of large $N$ gauge theory accessible by numerical simulation. 

This paper is organized as follows. 
In section 2, we explain the reduction of Wilson's lattice gauge theory 
to the one-site model in the large $N$ limit. 
In section 3, we study the four-dimensional case.
In addition to the standard plaquette action, 
we put the adjoint term in the action 
and study the system with 
the two coupling constants. 
%We explain a simple property of this system due to the factorization, 
%valid in the large $N$ limit, 
%which plays an essential role in our subsequent argument. 
In section 5, we show our results for the six-dimensional case. 
Section 6 is devoted to conclusion and discussion.

\section{Reduction to one-site model}

In Wilson's lattice gauge theory, SU($N$) gauge theory can be defined 
with the action 
\beq
S=- N \beta \sum_{n} \sum_{\mu\neq\nu}
\mbox{tr} (U_{n,\mu} U_{n+\mu,\nu} 
U_{n+\nu,\mu}^{\dagger} U_{n,\nu}^{\dagger}).
\eeq
$U_{n,\mu}$'s are SU($N$) matrices and are called link variables. 
A gauge invariant observable can be given by the Wilson loop
\beq
W(C)= \frac{1}{N}
\mbox{tr} (U_{n, \alpha}
U_{n+\hat{\alpha}, \beta}
U_{n+\hat{\alpha}+\hat{\beta}, \gamma}
\cdots
U_{n-\hat{\omega}, \omega}),
\label{eq:wilsonloop}
\eeq
which is the trace of the ordered product of link variables 
along the loop $C$.

The one-site model which is equivalent to the above 
Wilson's theory can be defined with the action
\beq
S=- N \beta \sum_{\mu\neq\nu}
Z_{\mu\nu} \mbox{tr} (U_{\mu} U_{\nu} U_{\mu}^{\dagger} U_{\nu}^{\dagger}),
\label{eq:TEKaction}
\eeq
where $Z_{\mu\nu}$ is an element of the center group $Z_N$ 
and is called `twist'. 
A successful choice of the twist is given by 
%an element of the center group $Z_{N}$, 
%which should be properly chosen so that the equivalence holds true.
% , which is called `twist', 
\beqa
Z_{\mu\nu} &=& \exp (2\pi i/L)~~~~~~~~~~~~~\mbox{for}~~\mu < \nu, \\
Z_{\nu\mu} &=& Z_{\mu\nu}^{*},
\eeqa
where $L$ is an integer defined through
\beq
N=L^{D/2}
\label{eq:twist}
\eeq
with the space-time dimension $D$, which is taken to be even
\footnote{
For odd $D$, one can apply the reduction scheme only to the 
$(D-1)$ dimensions leaving one dimension unreduced. 
We prefer to restrict ourselves to even dimensions, 
{\it i.e.} $D=4$ and $D=6$ in order to avoid any technical complications. 
}.
The twist has been introduced by the authors of Ref. \cite{GAO} in order to 
cure a problem in the original Eguchi-Kawai model \cite{EguchiKawai}, 
where $Z_{\mu\nu}$'s were taken to be unity.
The model is thus called twisted Eguchi-Kawai model.
For a general review on twisted Eguchi-Kawai models, we refer the reader
to Ref. \cite{Das}.

The observable corresponding to the Wilson loop 
(\ref{eq:wilsonloop}) can be defined 
in the twisted Eguchi-Kawai model as
\beq
w(C) = \frac{1}{N}
(\prod_{\mu \neq \nu} Z_{\mu\nu}^{N_{P_{\mu\nu}}})
\mbox{tr} (U_{\alpha}U_{\beta}U_{\gamma}\cdots U_{\omega}),
\eeq
where $N_{P_{\mu\nu}}$ is the number of plaquettes in the 
($\mu$,$\nu$) direction on the minimal surface spanning the loop $C$.

Under the assumption of the factorization
\beq
\langle W(C_1) W(C_2) \cdots W(C_k) 
\rangle
=
\langle W(C_1) \rangle 
\langle W(C_2) \rangle \cdots
\langle W(C_k) \rangle 
+ O \left(\frac{1}{N^2} \right),
\eeq
which is valid generically in large $N$ gauge theory,
one can show that
$\langle w(C) \rangle$ calculated in the twisted Eguchi-Kawai model 
is equal to 
$\langle W(C) \rangle$ calculated in the original Wilson's theory 
on the $L^D$ lattice, where $L$ is related to $N$ through
eq. (\ref{eq:twist}).

In this sense, 
the twisted Eguchi-Kawai model is equivalent to the original Wilson's 
theory on the infinite lattice, in the large $N$ limit.
As is indicated in the above statement, the finite $N$ effects 
in the twisted Eguchi-Kawai model appear in two ways.
One is the violation of the factorization
which is the assumption of the equivalence, 
and the other is the finite $L$ effects in the corresponding 
Wilson's theory to which the one-site model is equivalent.

Let us now explain how to perform 
Monte Carlo simulation of the twisted Eguchi-Kawai model. 
In contrast to the ordinary Wilson's theory, the action (\ref{eq:TEKaction}) 
is not linear in terms of each link variable.
In order to make the heat bath algorithm 
\cite{heatbath} applicable to the model, 
we use the technique proposed by Ref. \cite{FabriciusHaan}. 
The idea is to 
introduce an auxiliary field $Q_{\mu\nu}$($1\le\mu<\nu\le D$),
which is a general complex $N \times N$ matrix, with the following action. 
\beqan
S&=& N \beta \sum_{\mu<\nu} \tr Q_{\mu\nu}^{\dagger}Q_{\mu\nu} \\
  &~& - N \beta \sum_{\mu<\nu} \tr Q_{\mu\nu}^{\dagger} 
  (t_{\mu\nu}U_{\mu}U_{\nu}+ t_{\nu\mu}U_{\nu}U_{\mu}   )  \\
 &~&   - N \beta \sum_{\mu<\nu} \tr Q_{\mu\nu}  
     (t_{\mu\nu}^{\ast}U_{\nu}^{\dagger}U_{\mu}^{\dagger}
   + t_{\nu\mu}^{\ast} U_{\mu}^{\dagger}U_{\nu}^{\dagger}   ) ,
\eeqan
where $t_{\mu\nu}$ is the square root of $Z_{\mu\nu}$. 
%, namely, 
%\beqan
%t_{\mu\nu} &=& \ee ^{\pi i / L} ~~~~~~~~~~~~~~~(\mbox{for}~~\mu<\nu) \\
%t_{\nu\mu} &=& t_{\mu\nu}^{\ast}.
%\eeqan
Since this action is linear in terms of $U_{\mu}$, 
we can use the heat bath algorithm for the update of $U_{\mu}$.
We update $U_{\mu}$ by successively multiplying it by matrices
each belonging to the $N(N-1)/2$ SU(2) subgroups of the 
SU($N$) \cite{CabibboMarinari}.
%There are $N(N-1)/2$ of such SU(2) subgroups and 
%$[N/2]$ of them can be dealt with simultaneously.
After updating all the $U_{\mu}$'s in this way, 
we perform the update of $Q_{\mu\nu}$, which can be done 
with little cost by generating Gaussian variables. 
This defines the `one sweep' of our system. 
%We update $U_{\mu}$ by successive multiplication of elements 
%in SU(2) subgroups of SU($N$).
For further technical details of the algorithm, 
we refer the reader to Ref. \cite{FabriciusHaan}.

\section{Results for four-dimensional case}

We first study the model in four dimensions. 
We take $N=16$ which corresponds to $L=4$ in the equivalent Wilson's 
theory.
In fig. 1, we plot the mean plaquette defined by
\beq
P=\left\langle 
\frac{1}{N} \frac{1}{D(D-1)}
\sum_{\mu\ne\nu} Z_{\mu\nu} \tr(U_{\mu}U_{\nu}U_{\mu}^{\dagger}
U_{\nu}^{\dagger})
\right\rangle .
\label{eq:meanplaquette}
\eeq
Each point is an average over 1000 sweeps. 
%We have performed 1000 sweeps measuring the observable 
%every 50 sweeps at each $\beta$.
%Each point is an average over 1000 sweeps.
We also plot the leading terms of the strong- and weak-coupling expansion 
for the Wilson's theory. 
\beq
P \sim \left\{ 
\begin{array}{ll}
\beta  & \mbox{~~~~~(for small $\beta$)} \\
\frac{1}{2D\beta}  & \mbox{~~~~~(for large $\beta$)}.
\end{array}
\right.
\eeq
Our data are in good agreement with those of Ref. \cite{GAO}.
%A first order phase takes place at $\beta \simeq 0.35$.
%Near the critical point, we observe a flip-flop between the two 
%phase.
We observe a clear indication for a first order phase transition 
at $\beta \simeq 0.35$.
We should note that a first order phase transition is already 
present at finite $N$, for $N \ge 4$,
when one uses the standard plaquette action \cite{Creutz}.
This first order phase transition is considered as 
a lattice artifact of the plaquette action, and can be removed, 
for example, by using the modified action 
\footnote{There may be various types of modifications that can be used 
instead of (\ref{eq:modifiedaction}).
In Ref. \cite{IIY}, it is reported that the first order phase 
transition is absent when one uses the renormalization-group improved action.}
\beq
S= - N \beta \sum_{P} \tr U(P) - \frac{\beta_{\aaa}}{2}
\sum_{P}\tr_{\aaa} U(P),
\label{eq:modifiedaction}
\eeq
where $U(P)$ is the ordered product of link variables around 
the given oriented plaquette $P$. 
The second term is the trace of $U(P)$ taken in the adjoint representation, 
which can be rewritten as
\beq
\tr_{\aaa} U(P) = |\tr U(P)|^2 - 1.
\eeq
Indeed, in Ref. \cite{ACM}, Monte Carlo simulation 
of the above system in the ordinary Wilson's theory 
has been done for $4 \le N \le 8$ and 
it is shown that the line of first order phase transition 
terminates at a point with a sufficiently large negative 
$\beta_{\aaa}$.

We, therefore, study large $N$ gauge theory 
with this modified action.
The adjoint term can be incorporated into the 
twisted Eguchi-Kawai model with 
the action
\beq
S=- N \beta \sum_{\mu\neq\nu}
Z_{\mu\nu} \mbox{tr} (U_{\mu} U_{\nu} U_{\mu}^{\dagger} U_{\nu}^{\dagger})
- \frac{\beta_{\aaa}}{2} \sum_{\mu\ne\nu} 
| Z_{\mu\nu} \tr U_{\mu} U_{\nu}
U_{\mu}^{\dagger}U_{\nu}^{\dagger} |^2.
\eeq
The second term, which corresponds to the adjoint term, 
can be dealt with by Metropolis algorithm. 
In order to increase the acceptance rate, we rewrite the action as 
\beq
S=- N (\beta + \beta_{\aaa} C) \sum_{\mu\neq\nu}
Z_{\mu\nu} \mbox{tr} (U_{\mu} U_{\nu} U_{\mu}^{\dagger} U_{\nu}^{\dagger})
- \frac{\beta_{\aaa}}{2} \sum_{\mu\ne\nu}
 | Z_{\mu\nu} \tr U_{\mu} U_{\nu}
U_{\mu}^{\dagger}U_{\nu}^{\dagger} - N C |^2 ,
%+ \mbox{const.},
\eeq
taking 
the arbitrary real constant $C$ to be approximately 
equal to the mean plaquette defined by eq. (\ref{eq:meanplaquette}).

In fig. 2, we show the mean plaquette
as a function of $\beta$ for $\beta_{\aaa}=-0.3,-0.7,-1.0$.
Each point is an average over 1000 sweeps.
We replot the data for $\beta_{\aaa}=0$.
As $\beta_{\aaa}$ goes to a larger negative value, 
the gap at the transition point decreases, and finally disappears. 
%For $\beta_{\aaa}=-10.0$, the data show a continuous curve, indicating 
%the disappearance of the phase transition. 
This implies that the first order phase transition can 
be removed by introducing a sufficiently large negative $\beta_{\aaa}$ 
even in the large $N$ limit. 

It is known that, in the large $N$ limit, 
the theory with ($\beta$, $\beta_{\aaa}$) is equivalent to 
the theory with ($\beta '$, $\beta_{\aaa} '$), if 
\beq
\beta ' = \beta + (\beta_{\aaa} - \beta_{\aaa}')
P(\beta, \beta_{\aaa}),
\label{eq:adjoint}
\eeq
where $P(\beta,\beta_{\aaa})$ is the mean plaquette 
at the coupling constants ($\beta$, $\beta_{\aaa}$) 
\cite{MPS,DasKogut}.

Using this equivalence theorem, 
we can predict the results for $\beta_{\aaa}=0$ 
with the input of the data for $\beta_{\aaa}=-0.3,-0.7,-1.0$.
In fig. 3 we show the prediction from the data for 
$\beta_{\aaa}=-0.3,-0.7,-1.0$ together with the data for $\beta_{\aaa}=0$. 
Note that there are three predicted values
for each $\beta$ in the critical region.
The middle point is unstable and can never be seen in the simulation. 
The other two points are stable or metastable.
The metastable vacuum is stabilized as $N$ increases
due to the small tunneling probability 
$\sim\exp(-\mbox{const.}N^2)$.
%and it becomes well-defined in the large $N$ limit.
Each of the three points corresponds to a solution to 
the Schwinger-Dyson equation for the Wilson loops \cite{DasKogut}.
%since the tunneling probability from the false vacuum 
%to the true vacuum is suppressed as $\sim \exp(-N^2 \cdot \mbox{const.})$.
The `equivalence' of the systems with different $\beta_{\aaa}$'s should, 
therefore, be 
understood in the sense that the two systems have a common solution 
to the Schwinger-Dyson equation.
One can see that
our data clearly satisfies the equivalence theorem.

\section{Results for six-dimensional case}

Let us turn to the six-dimensional case. 
In fig. 4, we show the mean plaquette as a function of $\beta$ 
for $N=64$ ($L=4$) with the standard plaquette action ($\beta_{\aaa}=0$). 
Each point is an average over $300$ sweeps.
We see a strong hysteresis indicating a first order 
phase transition. 
Note that the thermal cycle is substantially larger than that in the 
four-dimensional case. 
Fig. 5 shows the result for $N=27$ ($L=3$) with $\beta_{\aaa}=-10.0$.
Each point is an average over 300 sweeps.
Here we see a striking difference from the four-dimensional case. 
We see that there are two different $\beta$'s that give the same 
value for the mean plaquette.
It is obvious from the relation (\ref{eq:adjoint})
that the phase transition will never become continuous,
%even if we take $\beta_{\aaa}\longrightarrow \infty$ limit. 
however large a negative value of $\beta_{\aaa}$ we may take.

Using the equivalence theorem in the previous section, 
we can examine the asymptotic behavior of the system in 
the large negative $\beta_{\aaa}$ region.
From the relation (\ref{eq:adjoint}), we have
\beq
P(\beta,\beta_{\aaa})
= P(\beta+(\beta_{\aaa}-\beta_{\aaa}')P(\beta,\beta_{\aaa}),\beta_{\aaa}').
\eeq
We fix $\beta$ and $\beta_{\aaa}$ and consider the asymptotic behavior 
of the system with 
$\beta ' = \beta+(\beta_{\aaa}-\beta_{\aaa}')P(\beta,\beta_{\aaa})$
and $\beta _{\aaa} '$ for $ \beta _{\aaa} ' \rightarrow - \infty$.
Differentiating the above equation with respect to $\beta$, we obtain
\beq
\left. \frac{\del P}{\del \beta}
\right | _{\beta ' ,\beta_{\aaa}'}
\sim \frac{1}{|\beta_{\aaa}'|}.
\eeq
%for $\beta_{\aaa}-\beta_{\aaa}'\gg 1$. 
Therefore, denoting the horizontal distance of the two nearly parallel 
data lines in fig. 5 as $\Delta \beta$, we can write the gap in 
the mean plaquette as
\beq
\Delta P = \Delta \beta \cdot 
\left. \frac{\del P}{\del \beta}
\right | _{\beta ' ,\beta_{\aaa}'}
\sim \frac{\Delta \beta}{| \beta_{\aaa}' | }.
\eeq
%\beqan
%\Delta P &=& \Delta \beta \cdot 
%\left. \frac{\del P}{\del \beta}
%\right | _{\beta ' ,\beta_{\aaa}'} \\
%&\sim& \frac{\Delta \beta}{| \beta_{\aaa}' | }.
%\eeqan
Also the fluctuation of the mean plaquette $P$ can be written as
\beq
\delta P = \sqrt{\langle P^2 \rangle- \langle P \rangle^2} 
         = \sqrt { \frac{1}{N^2}
\left. \frac{\del P}{\del \beta}
\right | _{\beta ' ,\beta_{\aaa}'} } 
\sim \frac{1}{\sqrt{| \beta_{\aaa}'|}} \frac{1}{N}.
\eeq
%\beqan
%\delta P &=& \sqrt{\langle P^2 \rangle- \langle P \rangle^2} \\
%         &=& \sqrt { \frac{1}{N^2}
%\left. \frac{\del P}{\del \beta}
%\right | _{\beta ' ,\beta_{\aaa}'} } \\
%&\sim& \frac{1}{\sqrt{| \beta_{\aaa}'|}} \frac{1}{N}.
%\eeqan
Thus the fluctuation $\delta P$ as well as 
the gap $\Delta P$ vanishes 
in the $\beta_{\aaa} ' \rightarrow - \infty $ limit. 
Note also that the gap $\Delta P$ vanishes faster than 
the fluctuation $\delta P$, so that the discontinuity, 
which is a clear signal of the first order phase transition, 
becomes difficult to observe in the actual simulation. 
These results naturally explain the observations of Ref. \cite{KNO}, 
where five-dimensional pure SU(2) lattice gauge theory has been studied. 
%They reported that the latent heat becomes invisible for 
%a sufficiently large negative $\beta_{\aaa}$, 
%but that the energy fluctuation also becomes small, 
%which prevent them from making a definite conclusion as to 
%whether the first order phase transition turns into second order.

In the large $N$ limit, by virtue of the equivalence theorem, 
% due to the factorization, 
we can conclude that the line of first order phase transition 
continues to 
($\beta,\beta_{\aaa}$)=($\infty,-\infty$). 
Considering that the data of Ref. \cite{KNO} are qualitatively the 
same as ours, we may naturally expect that the same is true for 
$2 \le N < \infty$. 

One may think of enlarging the coupling-constant space 
where one searches for a second order phase transition. 
For this purpose, we consider those additional terms in the action 
which can be obtained 
from (\ref{eq:modifiedaction})
by replacing plaquettes with 
three-dimensional loops which consist of six links.
Such terms can be incorporated into the twisted Eguchi-Kawai model 
by the following additional terms.
\beqan
&&-\gamma_{\aaa} \sum_{\mu<\lambda} \sum_{\nu\neq\mu,\lambda}
|Z_{\mu\nu}Z_{\nu\lambda}Z_{\lambda\mu}^{\ast} 
\tr (U_{\mu}U_{\nu}U_{\lambda}
U_{\mu}^{\dagger}U_{\nu}^{\dagger}U_{\lambda}^{\dagger}) - N \alpha|^2 \\
&&-\gamma_{\aaa} \sum_{\mu<\nu<\lambda}
|Z_{\mu\nu}^{\ast}Z_{\nu\lambda}^{\ast}Z_{\lambda\mu}^{\ast} 
\tr (U_{\mu}U_{\nu}^{\dagger}U_{\lambda}
U_{\mu}^{\dagger}U_{\nu}U_{\lambda}^{\dagger}) - N \alpha|^2.
\eeqan 
Corresponding to the mean plaquette, we define 
the mean three-dimensional loop by
\beqan
Q= \frac{1}{N}
\left[ 
\frac{2}{D(D-1)(D-2)}
\sum_{\mu<\lambda} \sum_{\nu\neq\mu,\lambda}
\mbox{Re} \{
Z_{\mu\nu}Z_{\nu\lambda}Z_{\lambda\mu}^{\ast} 
\tr (U_{\mu}U_{\nu}U_{\lambda}
U_{\mu}^{\dagger}U_{\nu}^{\dagger}U_{\lambda}^{\dagger})  \} \right.\\
\left. + \frac{6}{D(D-1)(D-2)}
\sum_{\mu<\nu<\lambda}
\mbox{Re} \{
Z_{\mu\nu}^{\ast}Z_{\nu\lambda}^{\ast}Z_{\lambda\mu}^{\ast} 
\tr (U_{\mu}U_{\nu}^{\dagger}U_{\lambda}
U_{\mu}^{\dagger}U_{\nu}U_{\lambda}^{\dagger})  \}  \right ].
\eeqan
%a further introduce 
%the following terms in the action. 
%\beq
%- N \gamma \sum_{Q} \tr U_{Q} - \frac{\gamma_{\aaa}}{2} 
%\sum_{Q} |\tr U_{Q}|^2,
%\eeq
%where $Q$ denotes a three-dimensional loop made by six links.
A similar argument as with $\beta$,$\beta_{\aaa}$ and $P$ applies 
to $\gamma$,$\gamma_{\aaa}$ and $Q$, where $\gamma = \gamma_{\aaa} \alpha$.

In fig. 6, we show the data for ($P$,$Q$) in the ordered phase 
and in the disordered phase 
for four sets of ($\beta$,$\gamma$)
with $\beta_{\aaa}=-1.0$ and $\gamma_{\aaa}=-2.0$.
Each point is an average over 50 sweeps.
Note that the quadrilaterals spanned by the data points
in the two phases have an overlap.
This means that the same $(P,Q)$ can be obtained 
for different $(\beta,\gamma)$'s with fixed 
$\beta_{\aaa}$ and $\gamma_{\aaa}$.
We can thus conclude that 
the phase transition will not become continuous
%there is no second order phase transition 
even in the enlarged coupling-constant space.

\vspace{1cm}

\section{Summary and Discussion}
%\setcounter{equation}{0}
%\hspace*{\parindent}

In this paper we study large $N$ gauge theory 
in six dimensions as well as in four dimensions. 
The phase diagram in the large negative $\beta_{\aaa}$ region 
has been clarified by making use of a simple property
of the system with the two coupling constants $\beta$ and $\beta_{\aaa}$
due to the large $N$.

In four dimensions the line of first order phase transition 
terminates at a sufficiently large negative $\beta_{\aaa}$ 
and the two phases are actually connected analytically. 
At the end point of the line, the first order phase transition 
becomes second order. 
What is the continuum theory that can be defined at this end point?
Since the phase transition is not a deconfining one, 
the string tension does not scale to zero at the end point.
This means that minimum surfaces dominate the summation over surfaces 
spanning the Wilson loops. 
Considering, for example, the two-point function of plaquettes,
dominant contributions come from 
tube-like surfaces connecting the two plaquettes.
This means that the continuum theory looks more like particle theory 
rather than string theory.
This is consistent with the claim for SU(3) gauge theory 
that the continuum theory defined at this end point is
(trivial) $\phi^4$ theory \cite{Heller}.

In six dimensions, the line of first order phase transition 
continues to infinity.
Moreover, we see that this is the case even if we 
enlarge the coupling-constant space to include 
three-dimensional loops in addition to plaquettes.
It is natural to expect that this pattern is repeated
for other types of loops.
Thus we consider that there is no nontrivial fixed point 
in large $N$ gauge theory in six dimensions.

In the context of string theory,
it is tempting to interpret our conclusion as a result of 
tachyon instability of bosonic strings, 
which may be cured by introducing space-time supersymmetry. 
It is, therefore, probable that we may find a nontrivial fixed point 
in six dimensional {\it supersymmetric} large $N$ gauge theory, 
though it will require much more effort since we have to 
deal with dynamical fermions.

\vspace{1cm}
{\Large\bf \noindent Acknowledgment} 
%\begin{center} \begin{large}
%Acknowledgements
%\end{large} \end{center}

~

\noindent 
%\begin{center} \begin{large}
%Acknowledgements
%\end{large} \end{center}
I would like to thank H. Kawai for stimulating discussion. 
I am also grateful to S.R. Das and Y. Iwasaki for helpful 
communications during this work.
This work was supported by 
National Laboratory for High Energy Physics (KEK) 
as KEK Supercomputer Project, 
and the calculations were carried out on Fujitsu VPP500 at KEK.

\newpage

\centerline{\Large Figure captions}
\bigskip
\noindent Fig. 1
The mean plaquette $P$ as a function of the coupling constant $\beta$ 
in four-dimensional SU(16) gauge theory with the standard plaquette 
action ($\beta_{\aaa}=0$). 
The dashed line represents the strong coupling expansion $P=\beta$, 
whereas the dash-dotted line represents the weak-coupling expansion 
$P=1/(8\beta)$. 

\bigskip
\noindent Fig. 2
The mean plaquette $P$ as a function of the coupling constant $\beta$ 
in four-dimensional SU(16) gauge theory with 
$\beta_{\aaa}=-0.3$ (triangles),$-0.7$ (squares),$-1.0$ (diamonds). 
We replot the data for
$\beta_{\aaa}=0$ (crosses) shown in fig. 1.

\bigskip
\noindent Fig. 3
The predicted values for the mean plaquette $P$ 
as a function of the coupling constant $\beta$ for 
$\beta_{\aaa}=0$
with the input of the data for $\beta_{\aaa}=-0.3$ (triangles),
$-0.7$ (squares),$-1.0$ (diamonds) shown in fig. 2.
We also plot the original data for
$\beta_{\aaa}=0$ (crosses).

\bigskip
\noindent Fig. 4
The mean plaquette $P$ as a function of the coupling constant $\beta$ 
in six-dimensional SU(64) gauge theory with 
the standard plaquette action ($\beta_{\aaa}=0$). 
The dashed line represents the strong coupling expansion $P=\beta$, 
whereas the dash-dotted line represents the weak-coupling expansion 
$P=1/(12\beta)$.

\bigskip
\noindent Fig. 5
The mean plaquette $P$ as a function of the coupling constant $\beta$ 
in six-dimensional SU(27) gauge theory with 
$\beta_{\aaa}=-10.0$. 
The triangles are the data in the disordered phase, whereas 
the squares are the data in the ordered phase.

\bigskip
\noindent Fig. 6
The mean plaquette $P$ and the mean three-dimensional loop $Q$
in six-dimensional SU(27) gauge theory
are plotted for four sets of ($\beta$,$\gamma$)
with $\beta_{\aaa}=-1.0$ and $\gamma_{\aaa}=-2.0$.
The triangles are the data in the disordered phase, whereas 
the squares are the data in the ordered phase.

\newpage

\ \vspace*{3cm}

\begin{center}
\begin{minipage}[t]{12cm}
\begin{center}
\leavevmode\psfig{file=energy4dN4.eps,width=12cm,angle=0}
%\leavevmode\psfig{file=energy6dN3_adj.eps,width=10cm,angle=270}
%angle=270
\end{center}
\setlength{\baselineskip}{5mm}
%{\footnotesize 
%Figure 1: 
%The mean plaquette $P$ as a function of the coupling constant $\beta$ 
%in four-dimensional SU(16) gauge theory with the standard plaquette 
%action ($\beta_{\aaa}=0$). 
%The dashed line represents the strong coupling expansion $P=\beta$, 
%whereas the dash-dotted line represents the weak-coupling expansion 
%$P=1/(8\beta)$. }
\end{minipage}
\end{center}

\newpage

\ \vspace*{3cm}
\begin{center}
\begin{minipage}[t]{12cm}
\begin{center}
\leavevmode\psfig{file=energy4dN4_adj.eps,width=12cm,angle=0}
%\leavevmode\psfig{file=energy6dN3_adj.eps,width=10cm,angle=270}
%angle=270
\end{center}
\setlength{\baselineskip}{5mm}
%{\footnotesize 
%Figure 1: 
%The mean plaquette $P$ as a function of the coupling constant $\beta$ 
%in four-dimensional SU(16) gauge theory with the standard plaquette 
%action ($\beta_{\aaa}=0$). 
%The dashed line represents the strong coupling expansion $P=\beta$, 
%whereas the dash-dotted line represents the weak-coupling expansion 
%$P=1/(8\beta)$. }
\end{minipage}
\end{center}

\newpage

\ \vspace*{3cm}
\begin{center}
\begin{minipage}[t]{12cm}
\begin{center}
\leavevmode\psfig{file=energy4dN4_equiv.eps,width=12cm,angle=0}
%\leavevmode\psfig{file=energy6dN3_adj.eps,width=10cm,angle=270}
%angle=270
\end{center}
\setlength{\baselineskip}{5mm}
%{\footnotesize 
%Figure 1: 
%The mean plaquette $P$ as a function of the coupling constant $\beta$ 
%in four-dimensional SU(16) gauge theory with the standard plaquette 
%action ($\beta_{\aaa}=0$). 
%The dashed line represents the strong coupling expansion $P=\beta$, 
%whereas the dash-dotted line represents the weak-coupling expansion 
%$P=1/(8\beta)$. }
\end{minipage}
\end{center}

\newpage

\ \vspace*{3cm}
\begin{center}
\begin{minipage}[t]{12cm}
\begin{center}
\leavevmode\psfig{file=energy6dN4.eps,width=12cm,angle=0}
%\leavevmode\psfig{file=energy6dN3_adj.eps,width=10cm,angle=270}
%angle=270
\end{center}
\setlength{\baselineskip}{5mm}
%{\footnotesize 
%Figure 1: 
%The mean plaquette $P$ as a function of the coupling constant $\beta$ 
%in four-dimensional SU(16) gauge theory with the standard plaquette 
%action ($\beta_{\aaa}=0$). 
%The dashed line represents the strong coupling expansion $P=\beta$, 
%whereas the dash-dotted line represents the weak-coupling expansion 
%$P=1/(8\beta)$. }
\end{minipage}
\end{center}

\newpage

\ \vspace*{3cm}
\begin{center}
\begin{minipage}[t]{12cm}
\begin{center}
\leavevmode\psfig{file=energy6dN3_adj.eps,width=12cm,angle=0}
%\leavevmode\psfig{file=energy6dN3_adj.eps,width=10cm,angle=270}
%angle=270
\end{center}
\setlength{\baselineskip}{5mm}
%{\footnotesize 
%Figure 1: 
%The mean plaquette $P$ as a function of the coupling constant $\beta$ 
%in four-dimensional SU(16) gauge theory with the standard plaquette 
%action ($\beta_{\aaa}=0$). 
%The dashed line represents the strong coupling expansion $P=\beta$, 
%whereas the dash-dotted line represents the weak-coupling expansion 
%$P=1/(8\beta)$. }
\end{minipage}
\end{center}

\newpage

\ \vspace*{3cm}
\begin{center}
\begin{minipage}[t]{12cm}
\begin{center}
\leavevmode\psfig{file=energy3DL.eps,width=12cm,angle=0}
%\leavevmode\psfig{file=energy6dN3_adj.eps,width=10cm,angle=270}
%angle=270
\end{center}
\setlength{\baselineskip}{5mm}
%{\footnotesize 
%Figure 1: 
%The mean plaquette $P$ as a function of the coupling constant $\beta$ 
%in four-dimensional SU(16) gauge theory with the standard plaquette 
%action ($\beta_{\aaa}=0$). 
%The dashed line represents the strong coupling expansion $P=\beta$, 
%whereas the dash-dotted line represents the weak-coupling expansion 
%$P=1/(8\beta)$. }
\end{minipage}
\end{center}

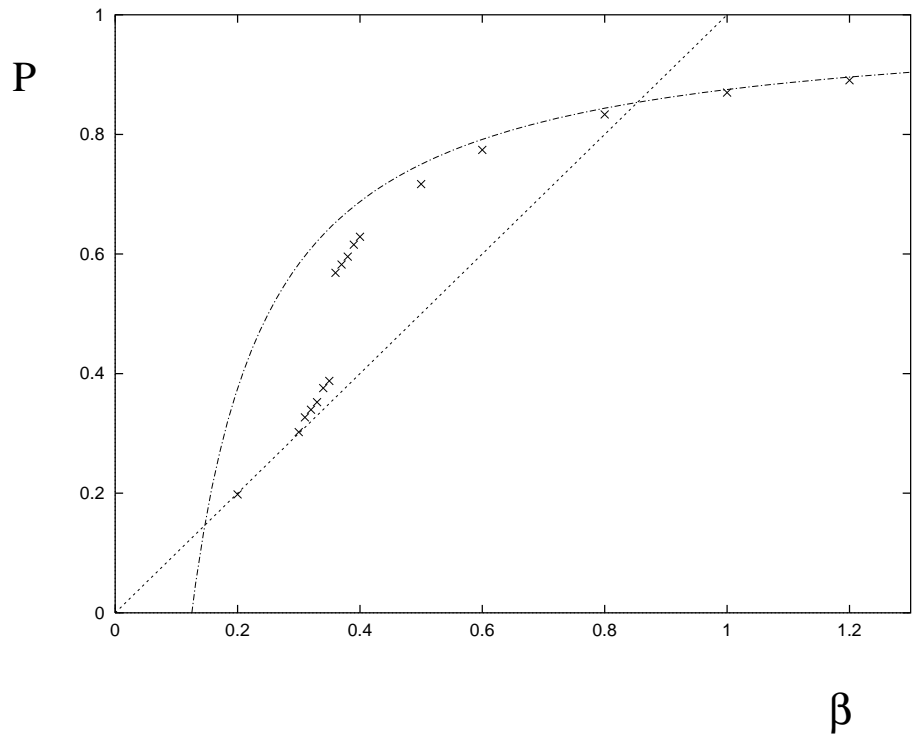
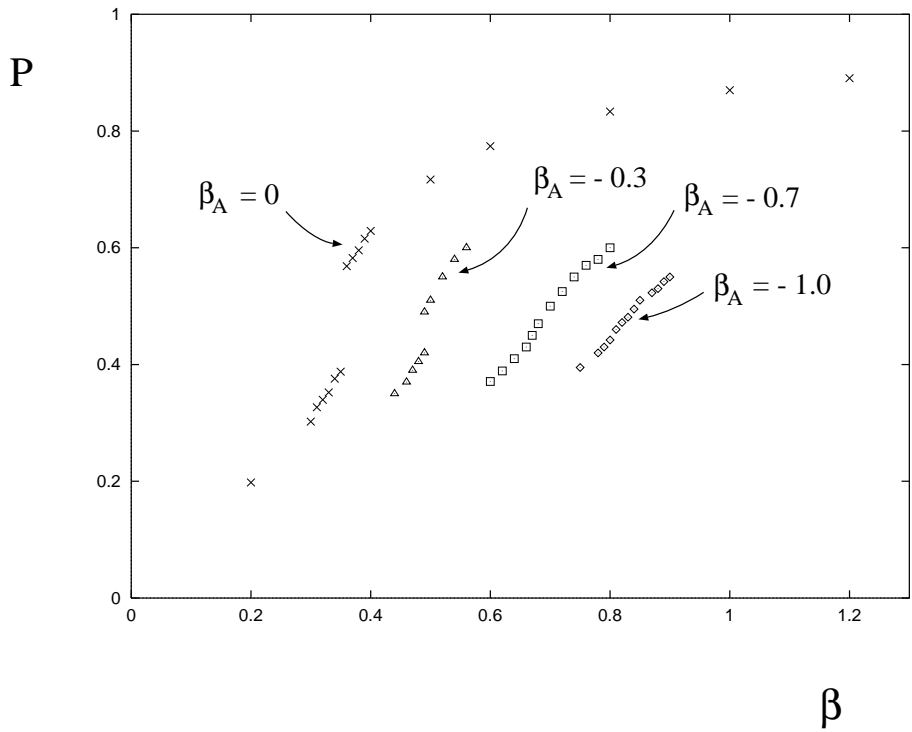
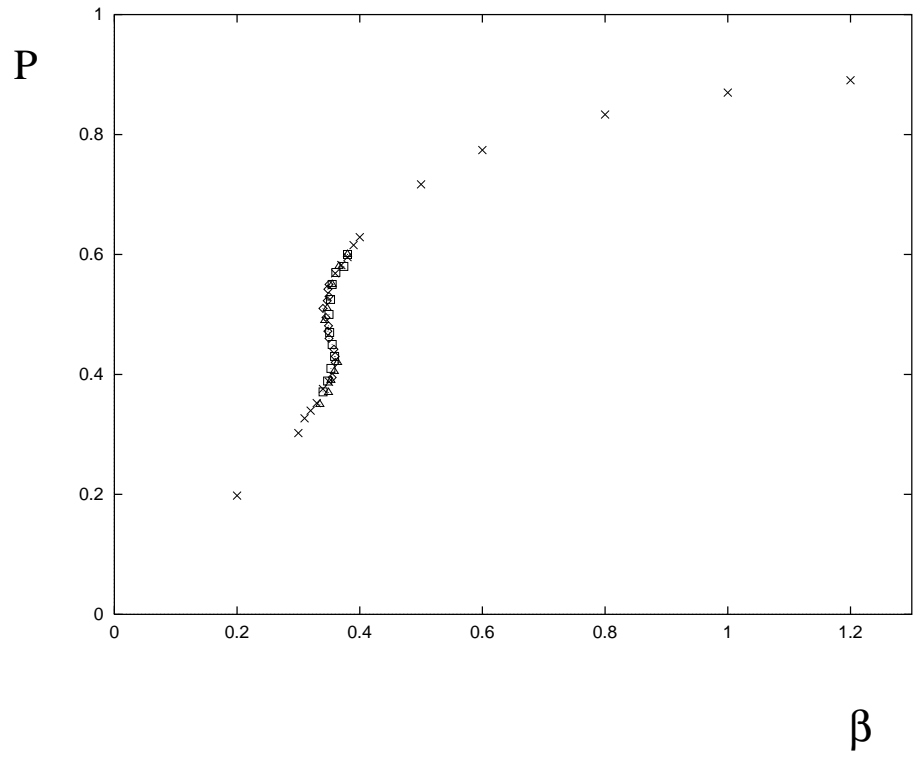
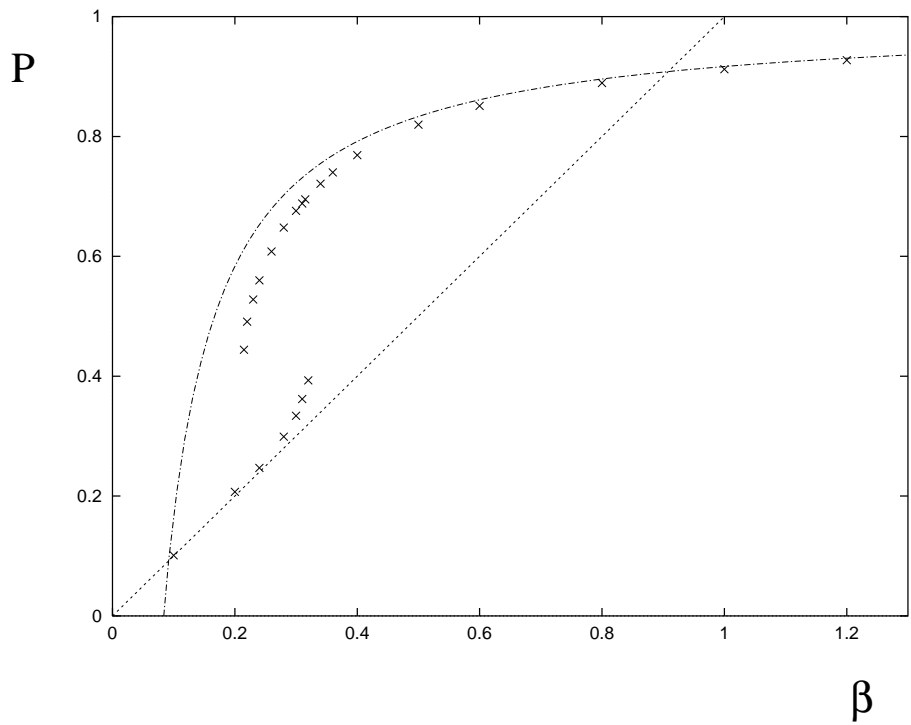
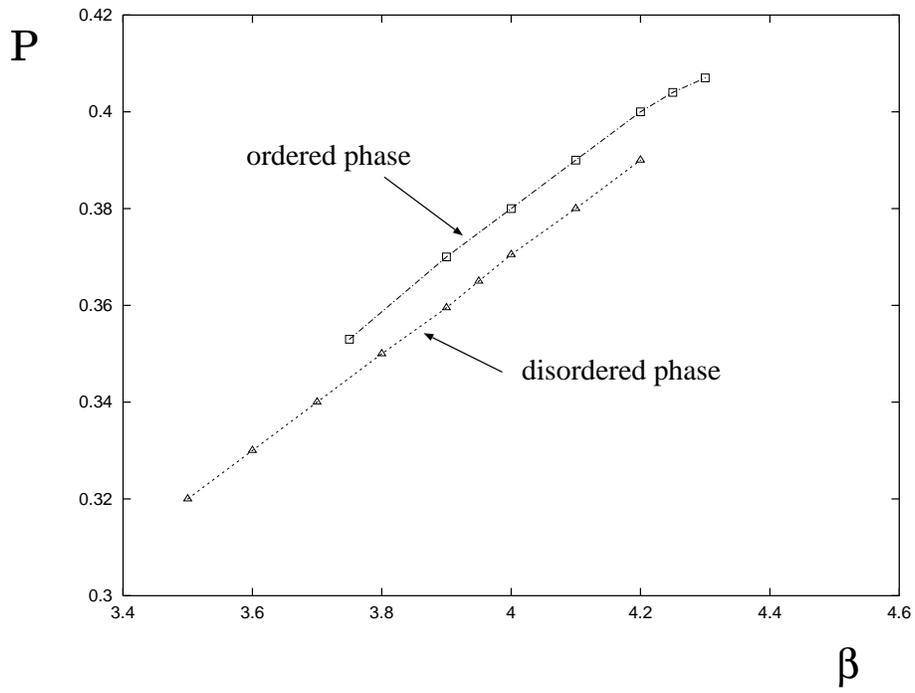
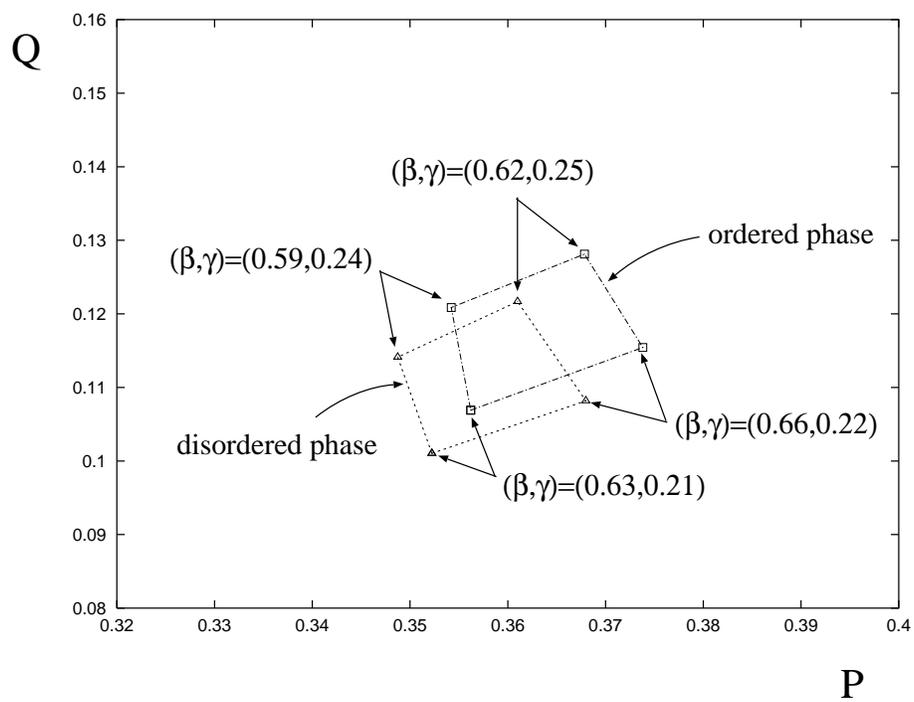
\end{document}